\documentclass[twocolumn,showpacs,preprintnumbers,amsmath,amssymb]{revtex4}

\usepackage{graphicx}
\usepackage{dcolumn}
\usepackage{bm}

\begin{document}

\preprint{}

\title{Strong spin-orbit interaction of light on the surface of atomically thin crystals}
\author{Mengxia Liu}
\author{Liang Cai}
\author{Shizhen Chen}
\author{Yachao Liu}
\author{Hailu Luo}\email{hailuluo@hnu.edu.cn}
\author{Shuangchun Wen}
\affiliation{Laboratory for Spin Photonics, School of
 Physics and Electronics, Hunan University, Changsha 410082,China}
\date{\today}

\begin{abstract}
The photonic spin Hall effect (SHE) can be regarded as a direct optical
analogy of the SHE in electronic systems where a refractive index gradient plays the role of electric potential.
However, it has been demonstrated that the effective refractive index
fails to adequately explain the light-matter interaction in atomically thin crystals.
In this paper, we examine the spin-orbit interaction on the surface of the freestanding atomically thin crystals.
We find that it is not necessary to involve the effective refractive index to describe the
spin-orbit interaction and the photonic SHE in the atomically thin crystals.
The strong spin-orbit interaction and giant photonic SHE have been predicted, which can be
explained as the large polarization rotation of plane-wave components in order to satisfy the transversality of photon.
\end{abstract}


\maketitle

\section{Introduction}\label{SecI}
Two-dimensional (2D) atomic crystals have extraordinary electronic and photonic properties which hold great promise in the application of photonics
and optoelectronics~\cite{Novoselov2004,Novoselov2005,Bonaccorso2010}. A fundamental understanding of the light-matter interaction in the 2D atomically thin crystals is therefore essential to optoelectronics applications.
Reflection and refraction are most common optical phenomena, which are governed by the boundary condition~\cite{Jackson1999}.
In general, the interpretation of reflection and refraction on the surface of 2D atomically thin crystals is treated as a homogeneous medium with an
effective refractive index and an effective thickness~\cite{Blake2007,Bruna2009,Kravets2010,Peters2011,Zhou2012,Golla2013}.
Recently, it has been demonstrated that the Fresnel model based on the certain thickness and effective refractive index
fails to explain the overall experiments on light-matter interaction~\cite{Merano2016I,Merano2016II,Merano2016III}.
However, the Fresnel model based on the zero-thickness interface
can give a complete and convincing description of all the experimental observation.
Here, the 2D atomic crystals can be regarded as zero-thickness interface (a real 2D system).

As a fundamental physical effect in light-matter interaction,
spin-orbit coupling of light is attributed to the transverse nature of the photonic polarization.
Photonic spin Hall effect (SHE) manifesting itself as spin-
dependent splitting in light-matter interaction is considered as a result of spin-orbit interaction of
light~\cite{Onoda2004,Bliokh2006,Hosten2008}. The photonic SHE can be regarded as a direct optical
analogy of the SHE in electronic systems~\cite{Dyakonov1971,Hirsch1999,Murakami2003,Sinova2004,Wunderlich2005} where the spin
electrons and electric potential are replaced by spin photons
and a refractive index gradient, respectively.
The analogy has been extensively demonstrated effective for the photonic SHE in 3D bulk
crystals~\cite{Bliokh2007,Bliokh2008,Aiello2008,Luo2009,Menard2010,Luo2011,Zhou2013,Korger2014,Ren2015}. However, the effective refractive index fails to adequately explain the light-matter interaction in 2D atomic crystals.
It would be interesting how to describe the spin-orbit interaction on the surface of 2D atomic crystals.

In this paper, we examine the spin-orbit coupling of light on the surface of the freestanding atomically thin crystals.
We develop a general model to describe the spin-orbit interaction of light on the surface of 2D atomic crystals.
We find that it is not necessary to involve the effective refractive index to describe the spin-orbit interaction and photonic SHE on the surface of atomically thin crystals. Based on this model,  the spin-dependent spatial and angular shifts in photonic SHE can
be obtained. The strong spin-orbit interaction and the giant photonic SHE have been predicted, which can be explained as the large polarization rotation of plane-wave components in order to satisfy the transversality of photon.

\section{A general model for spin-orbit interaction of light}\label{SecII}
\begin{figure}
\centerline{\includegraphics[width=8cm]{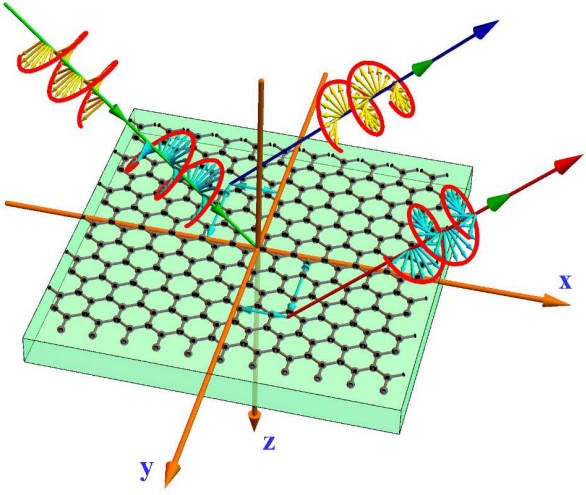}}
\caption{\label{Fig1} Schematic illustrating the photonic SHE
of wavepacket reflected on the surface of atomically thin crystal.
On the surface of 2D atomic crystal, the photonic SHE occurs which manifests as the spin-dependent splitting. For the freestanding atomically thin crystal, we can choose the refractive index of substrate as $n=\sqrt{\varepsilon/\varepsilon_0}=1$.}
\end{figure}

We first establish a general model to describe the spin-orbit interaction on the surface of 2D atomic crystals. Let us consider a Gaussian wavepacket with monochromatic frequency $\omega$ impinging from air to the surface of the 2D atomic crystal as shown in Fig.~\ref{Fig1}.
The $z$ axis of the laboratory Cartesian frame ($x,y,z$) is normal to the surface of the 2D atomic crystal.
A sheet of 2D atomic crystal is placed on the top of a dielectric substrate.
In addition, the coordinate frames ($x_i,y_i,z_i$) and
($x_r,y_r,z_r$) are used to denote central
wave vector of incidence and reflection, respectively.

In order to keep the discussion as general
as possible, the conductivity and susceptibility tensors for the 2D atomic crystals can be written as
\begin{eqnarray}
\sigma_T=\left(\begin{array}{lcr}\sigma_{pp} & \sigma_{ps}\\\sigma_{sp}&\sigma_{ss}\end{array}\right),~~~
\chi_T=\left(\begin{array}{lcr}\chi_{pp} & \chi_{ps}\\\chi_{sp}&\chi_{ss}\end{array}\right)
.\label{sigmachi}
\end{eqnarray}
The conductivity and susceptibility tensors can be applied to describe different 2D atomic crystals, such as graphene~\cite{Novoselov2004}, boron-nitride~\cite{Geim2013}, and black phosphorus~\cite{FXia2014}.

Based on the boundary conditions, the incident, reflected, and transmitted amplitudes satisfy the following equations:
\begin{eqnarray}
E^s_i+E^s_r=E^s_t\label{ReflectI},
\end{eqnarray}
\begin{eqnarray}
\cos\theta_i(E^p_i-E^p_r)=\cos\theta_tE^p_t\label{ReflectII},
\end{eqnarray}
\begin{eqnarray}
\frac{\cos\theta_i}{Z_0}(E^s_i-E^s_r)&=&\left(\sigma_{ss}+\frac{ik\chi_{ss}}{Z_0} +\frac{\cos\theta_t}{Z}\right)E^s_t\nonumber\\
&&+\left(\frac{ik\chi_{sp}}{Z_0}+\sigma_{sp}\right)\cos\theta_tE^p_t\label{ReflectIII},
\end{eqnarray}
\begin{eqnarray}
\frac{1}{Z_0}(E^p_i+E^p_r)&=&\left(\sigma_{pp}\cos\theta_t+\frac{ik\chi_{pp}}{Z_0}\cos\theta_t +\frac{1}{Z}\right)E^p_t\nonumber\\
&&+\left(\frac{ik\chi_{ps}}{Z_0}+\sigma_{ps}\right)E^s_t\label{ReflectIV}.
\end{eqnarray}
Here, $p$ and $s$ represent the parallel and perpendicular
polarization states, respectively. $\theta_i$ is the angle of incidence, $\theta_t$ is the transmission angle. ${Z_0}$ is the impedance in air and ${Z}$ is the impedance in media.
The Fresnel's coefficients are determined by the incident and reflected
amplitudes: $r_{pp}=E^p_r/E^p_i$ , $r_{ss}=E^s_r/E^s_i$, $r_{ps}=E^p_r/E^s_i$ and
$r_{sp}=E^s_r/E^p_i$. From Eqs.~(\ref{ReflectI})-(\ref{ReflectIV}), the Fresnel's coefficients are obtained as
 \begin{equation}
  r_{pp}=\frac{\alpha^T_+\alpha_-^L+\beta}{\alpha_+^T\alpha_+^L+\beta}\label{RPP},
 \end{equation}
 \begin{equation}
  r_{ss}=-\frac{\alpha^T_-\alpha_+^L+\beta}{\alpha^T_+\alpha_+^L+\beta}\label{RSS},
 \end{equation}
\begin{equation}
r_{ps}=-r_{sp}=\frac{\Lambda}{\alpha^T_+\alpha_+^L+\beta}\label{Re-co}.
\end{equation}
Here, $\alpha^L_\pm=(k_{iz}\varepsilon\pm k_{tz}\varepsilon_0+i\varepsilon_0k_{iz}k_{tz}\chi_{pp}+k_{iz}k_{tz}\sigma_{pp}/\omega)/\varepsilon_0$,
  $\alpha^T_\pm=k_{tz}\pm k_{iz}+ik^2\chi_{ss}+\omega\mu_0\sigma_{ss}$,
  $\beta=-[ik_{iz}k_{tz}\chi_{ps}+k_{iz}k_{tz}\sigma_{ps}/(\omega\varepsilon_0)](ik^2\chi_{ps}+\omega\mu_0\sigma_{ps})/\mu_0$,
  $\Lambda=2k_{iz}k_{tz}(ik\chi_{ps}+Z_0\sigma_{ps})$,
$k_{iz}=k_i\cos\theta_i$, and $k_{tz}=k_t \cos\theta_t$; $\varepsilon_0$ , $\mu_0$  are
  permittivity and permeability in vacuum; $\varepsilon$ is the permittivity of substrate;
  $\sigma_{pp}$, $\sigma_{ss}$ and $\sigma_{ps}$ ($\sigma_{sp}$ )
  denote the longitudinal, transverse, and crossing-conductance  conductivity, respectively.

For horizontal polarization state $|{H}\rangle$ and vertical polarization state $|{V}\rangle$,
the reflected polarization states related to the incident polarization states can be written as
$[|{{H}}({k}_r)\rangle~|{{V}}({k}_r)\rangle]^T={M}_{R}[|{H}({k}_i)\rangle~|{V}({k}_i)\rangle]^T$.
Here, ${M}_{R}$ can be expressed as
\begin{eqnarray}
\left[
\begin{array}{cc}
r_{pp}-\frac{2k_{ry}\cot\theta r_{ps}}{k_{0}} &r_{ps}+\frac{k_{ry}\cot\theta (r_{pp}+r_{ss})}{k_{0}} \\
r_{sp}-\frac{k_{ry}\cot\theta (r_{pp}+r_{ss})}{k_{0}} &
r_{ss}-\frac{2k_{ry}\cot\theta r_{ps}}{k_{0}}
\end{array}\right]\label{Matrix},
\end{eqnarray}
where $k_0=\omega/c$ is the wavevector in vacuum. In above equation, the boundary condition
$k_{rx}=-k_{ix}$ and $k_{ry}= k_{iy}$ have been introduced. The polarizations associated
with the angular spectrum components experience different rotations
in order to satisfy the boundary condition after reflection.

In the spin basis set, the polarization states of $|{H}\rangle$ and $|{V}\rangle$ can be decomposed into two orthogonal spin components
$|{H}\rangle=(|{+}\rangle+|{-}\rangle)$, and
$|{V}\rangle=i(|\mathbf{-}\rangle-|{+}\rangle)/\sqrt{2}$,
where $|{+}\rangle$ and $ |{-}\rangle$ represent the left- and right-circular polarization components, respectively.
We assume that the wavefunction in momentum space can be specified by the following expression
\begin{equation}
|\Phi\rangle=\frac{w_{0}}{\sqrt{2\pi}}\exp\left[-\frac{w^{2}_{0}(k_{ix}^{2}+k_{iy}^{2})}{4}\right]\label{GaussianWF},
\end{equation}
where $w_{0}$ is the width of wave function. The total wave function is made up of the packet spatial extent and
the polarization state. From Eqs.~(\ref{Matrix}) and (\ref{GaussianWF}), the reflected wave function
 $|{\psi}^{H}_{r}\rangle$ and $|{\psi}^{V}_{r}\rangle$ in the momentum space can be obtained as
 \begin{eqnarray}
  |{\psi}^{H}_{r}\rangle&=&\frac{r_{pp}{\pm}ir_{ps}}{\sqrt{2}}(1{\mp}ik_{rx}\delta_{x\pm}^H{\pm}ik_{ry}\delta_{y\pm}^H)\nonumber\\
 &&\times\exp\left[-\frac{w^{2}_{0}(k_{ix}^{2}+k_{iy}^{2})}{4}\right]|\pm\rangle,\label{WFH}
  \end{eqnarray}
  \begin{eqnarray}
   |{\psi}^{V}_{r}\rangle&=&\frac{r_{ps}{\mp}ir_{ss}}{\sqrt{2}}(1{\pm}ik_{rx}\delta_{x\pm}^V{\pm}ik_{ry}\delta_{y\pm}^V)\nonumber\\
   &&\times\exp\left[-\frac{w^{2}_{0}(k_{ix}^{2}+k_{iy}^{2})}{4}\right]|\pm\rangle,\label{WFV}
\nonumber\\\label{WFV}
  \end{eqnarray}
 Here, $\delta_{x\pm}^H= (\partial r_{ps}/\partial \theta_i)/(r_{pp}\pm ir_{ps})$, $\delta_{y\pm}^H=(r_{pp}+r_{ss})\cot\theta+\partial r_{ps}/\partial \theta]/(r_{pp}\pm ir_{ps})-2i\cot\theta r_{ps}/[k_0(r_{pp}\pm ir_{ps})]$, $\delta_{x\pm}^V= (\partial r_{ps}/\partial \theta)/(r_{ss}\pm ir_{ps})$, $\delta_{y\pm}^V=(r_{pp}+r_{ss})\cot\theta+\partial r_{ps}/\partial \theta]/[(r_{ss}\pm ir_{ps})-2i\cot\theta r_{ps}/[k_0(r_{ss}\pm ir_{ps})]$.
For weak spin-orbit interaction, $\delta^{H,V}_{rx}\ll{w_0}$ and $\delta^{H,V}_{ry}\ll{w_0}$, the reflected wavefunctions can be written as
\begin{eqnarray}
|{\psi}_r^{H}\rangle&\approx&\frac{r_{pp}\pm i r_{sp}}{\sqrt{2}}\exp(\mp ik_{rx}\delta_{rx\pm}^H\pm ik_{ry}\delta_{ry\pm}^H)\nonumber\\
&&\times\exp\left[-\frac{w^{2}_{0}(k_{ix}^{2}+k_{iy}^{2})}{4}\right]|\pm\rangle,\label{WPHI}
\end{eqnarray}
\begin{eqnarray}
|{\psi}_r^{V}\rangle&\approx&\frac{r_{ps}\mp ir_{ss}}{\sqrt{2}}\exp(\pm ik_{rx}\delta_{rx\pm}^V\pm ik_{ry}\delta_{ry\pm}^V)\nonumber\\
&&\times\exp\left[-\frac{w^{2}_{0}(k_{ix}^{2}+k_{iy}^{2})}{4}\right]|\pm\rangle.\label{WPHI}
\end{eqnarray}
Here, we have introduced the approximations: $1+i\sigma k_{rx}\delta^{H,V}_{rx\pm}\approx\exp(i\sigma k_{rx}\delta^{H,V}_{rx\pm})$ and $1+i\sigma k_{ry}\delta^{H,V}_{ry}\approx\exp(i\sigma k_{ry}\delta^{H,V}_{ry})$, where $\sigma$ being the Pauli operator.
The origin of the spin-orbit interaction
terms $\exp(i\sigma k_{rx}\delta^{H,V}_{rx})$ and $\exp(i\sigma k_{ry}\delta^{H,V}_{ry})$ lie in the transverse nature of the photon
polarization: The polarizations associated with
the plane-wave components experience different
rotations in order to satisfy the transversality
in reflection. In general, the phases  $\varphi_G=k_{rx}\delta^{H,V}_{rx}$ and
$\varphi_G=k_{ry}\delta^{H,V}_{ry}$ can be regarded as the spin-redirection Berry phases~\cite{Berry1984,Bliokh2015}.
It should be noted that the above approximations
do not hold for strong spin-orbit interaction $\delta^{H,V}_{rx}\approx{w_0}$ or $\delta^{H,V}_{ry}\approx{w_0}$ .

\section{Strong spin-orbit interaction}\label{SecII}
We now develop the theoretical mode to describe the strong spin-orbit interaction of light on the surface of
atomically thin crystals. Here, we restrict the isotropic case (such as graphene and boron-nitride),
where the Fresnel reflection coefficients $r_{ps}=r_{sp}=0$. By making use of Taylor
series expansion based on the arbitrary angular spectrum component,
$r_{pp}$ and $r_{ss}$ can be expanded as a polynomial of $k_{ix}$:
\begin{eqnarray}
r_{pp}&=&r_{pp}(k_{ix}=0)+k_{ix}\left[\frac{\partial
r_{pp}(k_{ix})}{\partial
k_{ix}}\right]_{k_{ix}=0}\label{Talorkx},
\end{eqnarray}
\begin{eqnarray}
r_{ss}&=&r_{ss}(k_{ix}=0)+k_{ix}\left[\frac{\partial
r_{ss}(k_{ix})}{\partial
k_{ix}}\right]_{k_{ix}=0}\label{Talorky}.
\end{eqnarray}
To accurately describe the strong spin-orbit interaction, the Fresnel
reflection coefficients are confined to the first order in Taylor series expansion.
We then obtain
 \begin{eqnarray}
  |{\psi}^{H}_{r\pm}\rangle&=&\bigg[r_{pp}-\frac{k_{rx}}{k_0}\frac{\partial r_{pp}}{\partial\theta_i}{\mp}i\frac{k_{ry}\cot\theta_i}{k_{0}}(r_{pp}+r_{ss})\nonumber\\
  &&{\mp}i\frac{k_{rx}k_{ry}\cot\theta_i}{k_{0}^2}\left(\frac{\partial r_{pp}}{\partial\theta_i}+\frac{\partial r_{ss}}{\partial\theta_i}\right)\bigg]\nonumber\\
  &&\times\exp\left[-\frac{w^{2}_{0}(k_{rx}^{2}+k_{ry}^{2})}{4}\right]|\pm\rangle
  \label{WFHSI}
  \end{eqnarray}
  \begin{eqnarray}
  |{\psi}^{V}_{r\pm}\rangle&=&\bigg[r_{ss}-\frac{k_{rx}}{k_0}\frac{\partial r_{ss}}{\partial\theta_i}{\mp}i\frac{k_{ry}\cot\theta_i}{k_{0}}(r_{pp}+r_{ss})\nonumber\\
  &&{\mp}i\frac{k_{rx}k_{ry}\cot\theta_i}{k_{0}^2}\left(\frac{\partial r_{pp}}{\partial\theta_i}+\frac{\partial r_{ss}}{\partial\theta_i}\right)\bigg]\nonumber\\
  &&\times\exp\left[-\frac{w^{2}_{0}(k_{ix}^{2}+k_{iy}^{2})}{4}\right]|\pm\rangle.\label{WFVSI}
  \end{eqnarray}
The large polarization in momentum space ($k$) will induces a giant spin-dependent splitting in position space.

\begin{figure}
\centerline{\includegraphics[width=8cm]{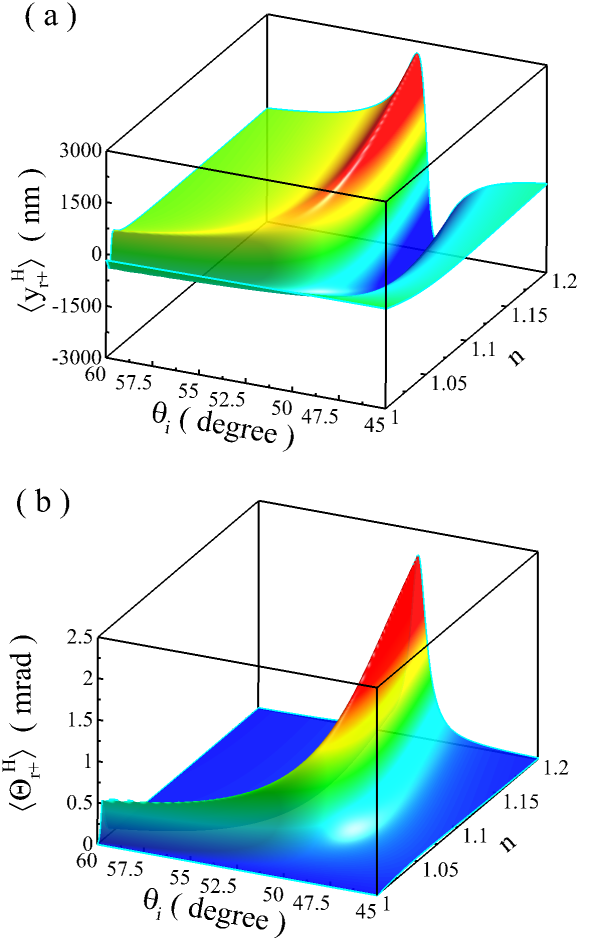}}
\caption{\label{Fig2} Strong spin-orbit interaction of light on the surface of 2D atomic crystals.
The spatial shift (a) and angular shift
(b) on the surface of 2D atomic crystals as a function
of incident angles $\theta_i$ and the refractive index of substrate $n$.
The 2D atomic crystal is chosen as single-layer graphene with $\sigma_{pp}=\sigma_{ss}=6.08\times10^{-5}\Omega$, $\sigma_{ps}=\sigma_{sp}=0$,
$\chi_{pp}=\chi_{ss}=1.0\times10^{-9}\mathrm{m}$, and $\chi_{ps}=\chi_{sp}=0$.}
\end{figure}

The transverse spatial and angular shifts of wave-packet at initial position ($z_r=0$) are given by
\begin{equation}
\langle{y_{r\pm}^{H,V}}\rangle=\frac{\langle\psi_{r\pm}^{H,V}|\partial_{k_{ry}}
|\psi_{r\pm}^{H,V}\rangle}{\langle\psi_{r\pm}^{H,V}|\psi_{r\pm}^{H,V}\rangle}\label{PYHV}.
\end{equation}
\begin{equation}
\langle{\Theta_{ry\pm}^{H,V}}\rangle=\frac{1}{k_0}\frac{\langle\psi_{r\pm}^{H,V}|{k_{ry}}
|\psi_{r\pm}^{H,V}\rangle}{\langle\psi_{r\pm}^{H,V}|\psi_{r\pm}^{H,V}\rangle}.\label{AYHV}
\end{equation}
Substituting Eqs.~(\ref{WFHSI}) and (\ref{WFVSI}) into Eqs.~(\ref{PYHV}) and (\ref{AYHV}), respectively,
the transverse spatial and angular shifts for two spin components are achieved.

Figure~\ref{Fig2} shows the transverse spatial and angular shifts
for the $|H\rangle$ polarization impinging on the surface of single-layer graphene.
The transverse shifts
are plotted as a function of incident angles and
the refractive index of substrate. For one-layer graphene at wavelength $633\mathrm{nm}$, the surface conductivity and the surface susceptibility values are chosen as $6.08\times10^{-5}\Omega$ and $\chi_{pp}=\chi_{ss}=1.0\times10^{-9}\mathrm{m}$, respectively~\cite{Merano2016I}.
The large spatial shifts occurs near a certain angle [Fig.~\ref{Fig2}(a)],
which can be regarded as the Brewster angle on reflection at the interface of air-substrate~\cite{Luo2011}.
The large angular shifts present due to the surface conductivity and the surface susceptibility
of the 2D atomic crystal [Fig.~\ref{Fig2}(b)]. It should be mentioned that there are no angular shifts present at the air-substrate interface.
The incident angle associated the large spatial and angular shifts increase with the decrease of the reflective index. 	
It should be interesting for the free standing 2D atomic crystals in vacuum where the refractive index of substrate is chosen as $n=1$.

\begin{figure}
\centerline{\includegraphics[width=8cm]{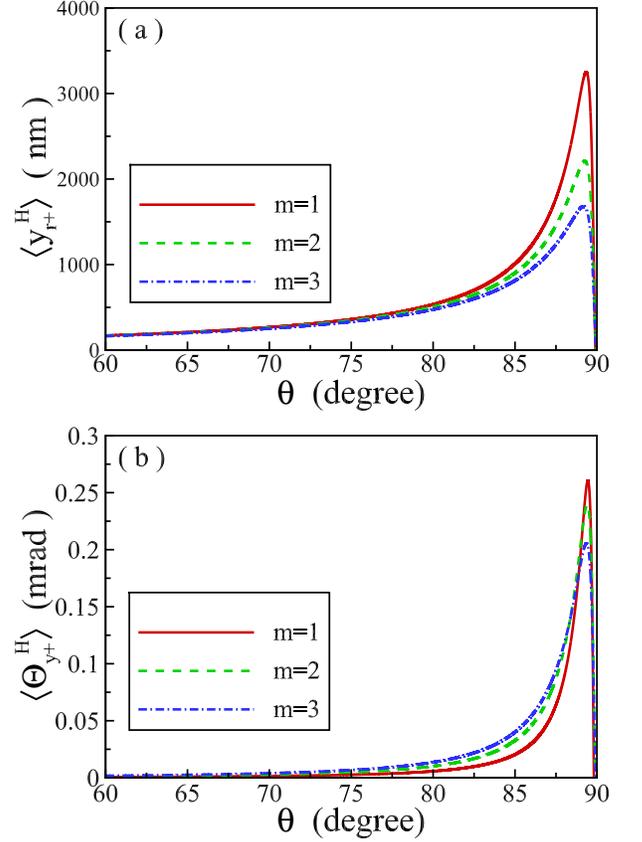}}
\caption{\label{Fig3}  Giant spin-dependent shifts in photonic SHE when the wave packet is reflected from the surface of freestanding
2D atomically thin crystal.
The spatial shifts (a) and the angular shifts (b) on the surface of atomically thin crystal with different layers $m=1,2,3$. The beam waist is chosen as $w_0=20\mathrm{{\mu}m}$.
Other parameters are the same as in Fig.~\ref{Fig2}.}
\end{figure}

 Assuming that an individual graphene sheet is a non-interacting monolayer, for few-layer graphene, the surface conductivity and the surface susceptibility values increase linearly with layer number $m$. The parameters for multi-layer graphene are obtained as $m\times6.08\times10^{-5}\Omega$ and $m\times1.0\times10^{-9}\mathrm{m}$. This assumption has been used to analyze the Goos-H\"{a}nchen effect on the surface of graphene, and the theoretic results coincide well with the experimental ones~\cite{Chen2017}. Figure~\ref{Fig3} shows the transverse spatial and angular shifts
for the polarization impinging on the surface of free-standing graphene with different layers. The obtained spatial shift reaches $3000\mathrm{nm}$ near the grazing angle, which is several times larger than the wave length [Fig.~\ref{Fig3}(a)].
Correspondingly, the angular shifts reaches $0.25\mathrm{mrad}$ [Fig.~\ref{Fig3}(b)]. Note that the large spatial and angular shifts in Goos-H\"{a}nchen effect have been predicted theoretically~\cite{Merano2016II}. In addition, the quantized beam shifts~\cite{Kamp2016,Cai2017} should also been enhanced in quantum Hall regime when the wavepacket reflection near the grazing angle.

We give a simple explanation of why the polarization rotation can be regarded as the
origin of photonic SHE. In general, an arbitrary linear polarization state can be decomposed into two orthogonally circular polarization states with opposite phases:
\begin{eqnarray}
\left(
\begin{array}{c}
\cos\gamma\\
\sin\gamma
\end{array}
\right)= \exp(+i\varphi_G)|\mathbf{+}\rangle+\exp(-i\varphi_G)|\mathbf{-}\rangle,\label{Jones}
\end{eqnarray}
where $\gamma$ is the polarization angle. The polarization rotation will induce a geometric phase gradient and ultimately lead to the spin-dependent shifts. When the polarization rotation occurs in momentum space, the spatial shift will be induced $\langle{y_{r\pm}}\rangle=\sigma\partial\varphi_G/\partial{k_{ry}}$.
Similarly, when the polarization rotation occurs in position space, the angular shift will be induced $\Delta{k_{ry\pm}}=\sigma\partial\varphi_G/\partial{y_r}$ and $\langle{\Theta_{ry\pm}}\rangle=\Delta{k_{ry\pm}}/k_r$.

\begin{figure}
\centerline{\includegraphics[width=8.5cm]{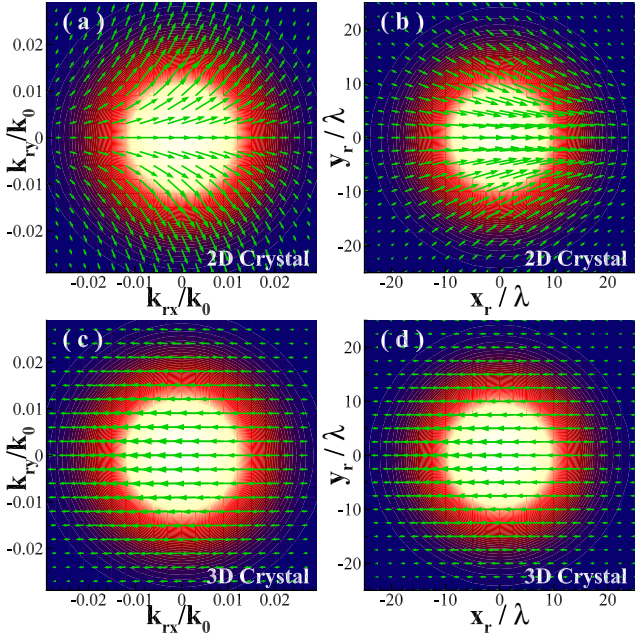}}
\caption{\label{Fig4}[(a) and (b)] Polarization rotation of the beam reflected on the surface of freestanding 2D atomic crystal without substrate.
[(c) and (d)] Polarization rotation on the surface of the 3D bulk crystal $n=1.515$.
The spin-orbit interaction can be explained as
the polarization rotation in momentum space and position space.
Left column: Polarization rotation in momentum
 space. Right column: Polarization rotation in position space.
 The incident angle is chosen as $\theta_{i}=85^{\circ}$ and the beam waist is chosen as $w_0=10\mathrm{{\mu}m}$.
 Other parameters are the same as in Fig.~\ref{Fig2}.
 To make the polarization rotation characteristics more noticeable, we amplify
 the rotation angles by $10$ times. }
\end{figure}

We now examine the polarization rotation characteristics of wave-packet after reflection. From Eqs.~(\ref{Matrix}) and (\ref{GaussianWF}), the representation of the reflected wave function can be written as
\begin{eqnarray}
|\psi_{r}^H\rangle&=&\exp\left[-\frac{w^{2}_{0}(k_{rx}^{2}+k_{ry}^{2})}{4}\right]\bigg[\left(r_{pp}-\frac{k_{rx}}{k_0}\frac{\partial r_{pp}}{\partial\theta_i}\right)|{H}\rangle\nonumber\\
 &&-\frac{k_{ry}\cot\theta_i}{k_{0}}(r_{pp}+r_{ss})|{V}\rangle+\frac{k_{rx}k_{ry}\cot\theta_i}{k_{0}^2}\nonumber\\
 &&\times\left(\frac{\partial r_{pp}}{\partial\theta_i}+\frac{\partial r_{ss}}{\partial\theta_i}\right)|{V}\rangle\bigg],\label{HKIC}
\end{eqnarray}
\begin{eqnarray}
|\psi_{r}^V\rangle&=&\exp\left[-\frac{w^{2}_{0}(k_{rx}^{2}+k_{ry}^{2})}{4}\right]\bigg[\left(r_{ss}-\frac{k_{rx}}{k_0}\frac{\partial r_{ss}}{\partial\theta_i}\right)|{V}\rangle\nonumber\\
 &&+\frac{k_{ry}\cot\theta_i}{k_{0}}(r_{pp}+r_{ss})|{H}\rangle-\frac{k_{rx}k_{ry}\cot\theta_i}{k_{0}^2}\nonumber\\
 &&\times\left(\frac{\partial r_{pp}}{\partial\theta_i}+\frac{\partial r_{ss}}{\partial\theta_i}\right)|{H}\rangle\bigg],\label{VKIC}
\end{eqnarray}
The wave function in position space is the Fourier transform of the wave function in momentum space:
\begin{equation}
|\Phi_{r}^{H,V}\rangle=\int\int{dk_{rx}dk_{ry}}|\psi_{r\pm}^{H,V}\rangle|k_{rx},k_{ry}\rangle.\label{Fourier}
\end{equation}
In fact, after the angular spectrum of incident wave function is known, Eq.~(\ref{Fourier})
together with Eqs.~(\ref{HKIC}) and (\ref{VKIC}) provides the general
representation of reflected wave function in position space:
\begin{eqnarray}
|\Phi_{r}^H\rangle&=&\exp\left[-\frac{(x_{r}^{2}+y_{r}^{2})}{w^{2}_{0}}\right]\bigg[\left(r_{pp}-\frac{ix_r}{z_R}\right)|{H}\rangle\nonumber\\
&&-\frac{iy_r}{z_R}\cot\theta_i(r_{pp}+r_{ss})|{V}\rangle-\frac{x_ry_r}{z_R^2}\cot\theta_i\nonumber\\
&&\times\left(\frac{\partial
r_{pp}}{\partial\theta_i}+\frac{\partial
r_{ss}}{\partial\theta_i}\right)\bigg]|{V}\rangle\bigg]\label{HPR},
\end{eqnarray}
\begin{eqnarray}
|\Phi_{r}^V\rangle&=&\exp\left[-\frac{(x_{r}^{2}+y_{r}^{2})}{w^{2}_{0}}\right]\bigg[\left(r_{ss}-\frac{ix_r}{z_R}\right)|{V}\rangle\nonumber\\
&&+\frac{iy_r}{z_R}\cot\theta_i(r_{pp}+r_{ss})|{H}\rangle+\frac{x_ry_r}{z_R^2}\cot\theta_i\nonumber\\
&&\times\left(\frac{\partial
r_{pp}}{\partial\theta_i}+\frac{\partial
r_{ss}}{\partial\theta_i}\right)|{H}\rangle\bigg]\label{VPR}.
\end{eqnarray}
The above expressions are only confine to the isotropic case. For anisotropic 2D atomic crystals,
more complex characteristics of polarization rotation would be involved.

We plot the polarization distributions of the reflected field in Fig.~\ref{Fig4}.
 In the reflection on the surface of the 2D atomic crystal, a large polarization rotation present in both position and momentum spaces [Fig.~\ref{Fig4}(a) and~\ref{Fig4}(b)]. Therefore, the large geometric phase gradient and giant spin-dependent splitting should also occur in momentum space and position space.
 As a comparison, the polarization rotation on the 3D bulk crystal are also plotted
 [Fig.~\ref{Fig4}(c) and~\ref{Fig4}(d)]. Interestingly, only a tiny polarization rotation appears in momentum space which ultimately induces a tiny spin-dependent splitting in position space.
 There are no polarization rotation appears in position space, and thereby no angular shift occurs in position space. Therefore, the spin-dependent splitting in position space is related to the
 polarization rotation in momentums space, while the splitting in momentum space is attributed to the polarization rotation in position space.

\section{Conclusions}
In conclusion, we have developed a general model to describe the spin-orbit interaction of light on the
surface of the free standing atomically thin crystals. In this model, the 2D atomic crystals can be regarded as zero-thickness interface. We have found that it is not necessary to involve the effective refractive index to describe the spin-orbit interaction and the photonic SHE in the atomically thin crystals. The giant photonic SHE  manifesting itself as large spin-dependent splitting in both position and momentum space have been theoretically predicted.
This strong spin-orbit interaction can be
explained as the large polarization rotation of plane-wave components in order to satisfy the transversality of photons.
We believe that these results may provide insights into
the fundamental properties of spin-orbit interaction of
light in 2D atomic crystals.

\begin{acknowledgements}
This research was supported by the National Natural Science Foundation
of China (Grants Nos. 11274106 and 11474089).
\end{acknowledgements}

\end{document}